%% file: monroe16.tex
\documentclass[apj]{emulateapj} 
\usepackage{longtable}

\usepackage{graphicx}
\usepackage{ulem} 

\begin{document}

\newcommand{\mipsmu}{$24\mu$m}
\newcommand{\uvcol}{(F390W--F814W)}
\newcommand{\Rproj}{R$_{\rm proj}$}
\newcommand{\rhalf}{r$_{1/2}$}
\newcommand{\Mstar}{M$_{\star}$}
\newcommand{\Msun}{M$_{\odot}$}
\newcommand{\duvcol}{$\delta_{\rm col}(M_{\star})$}
\newcommand{\duvcolRproj}{ $\delta_{\rm col}(R_{\rm proj})$}

\title{SG1120-1202: Mass-Quenching as Tracked by UV Emission in the
  Group Environment at \MakeLowercase{z}$=0.37$}
  
\author{Jonathan T. Monroe}
\affil{Physics and Astronomy, Texas A\&M University, College Station, TX 77840, USA}
\email{jonathan.monroe@tamu.edu}
\author{Kim-Vy H. Tran}
\affil{Physics and Astronomy, Texas A\&M University, College Station, TX 77840, USA}
\and
\author{Anthony H. Gonzalez}
\affil{Astronomy Department, University of Florida, Gainesville, FL 32611, USA}

\begin{abstract}

We use the {\it Hubble Space Telescope} to obtain WFC3/F390W imaging of the
supergroup SG1120-1202 at $z=0.37$, mapping the UV emission of
138 spectroscopically confirmed members.  We measure total
\uvcol\ colors and visually classify the UV morphology of individual
galaxies as ``clumpy'' or ``smooth.''  Approximately 30\% of the
members have pockets of UV emission (clumpy) and we identify for the
first time in the group environment galaxies with UV morphologies
similar to the jellyfish galaxies observed in massive clusters.
We stack the clumpy UV members and measure a shallow internal color
gradient, which indicates unobscured star formation is occurring
throughout these galaxies.  We also stack the four galaxy groups and
measure a strong trend of decreasing UV emission with decreasing
projected group distance (\Rproj).  We find that the strong
correlation between decreasing UV emission and increasing stellar mass
can fully account for the observed trend in \uvcol--\Rproj, i.e.,
mass-quenching is the dominant mechanism for extinguishing UV emission
in group galaxies.  Our extensive multi-wavelength analysis of
SG1120-1202 indicates that stellar mass is the primary predictor of UV
emission, but that the increasing fraction of massive (red/smooth)
galaxies at \Rproj$\lesssim2R_{200}$ and existence of jellyfish candidates is due to the group environment.

\end{abstract}
 
\keywords{galaxies: clusters: individual (SG1120-1202) - galaxies:
  evolution - galaxies: starburst - galaxies: morphology}

\section{Introduction}

How galaxies evolve as a function of their local environment continues
to be a rich topic of exploration for both observational and theoretical
studies.  The origins of well-established trends such as the
increasing fraction of blue/star-forming/disk-dominated galaxies in
galaxy clusters at higher redshift
\citep[e.g.][]{Butcher1978,Dressler1980,Stanford1998,Cooper2010}
seem to be rooted in environmental processes.  However, dissecting the
relative importance of physical mechanisms such as ram pressure
stripping \citep{Gunn1972,Abadi1999}, galaxy harassment
\citep{Moore1998a}, tidal interactions \citep{Byrd1990}, and gas
starvation \citep{Larson1980,Bekki2002} remains a challenge.  

In the hierarchical formation model \citep{Peebles1970}, galaxy
clusters are built by merging smaller groups.  
Observations of nearby groups show they have more in common
with galaxy clusters than with the field population, i.e., higher
early-type fractions and lower mean star-formation rates than the
field \citep{Hashimoto1998,Zabludoff1998,Tran2001}, and observations
at $z\gtrsim0.2$ highlight the importance of the group environment in
transforming late to early-type galaxies
\citep{Gallazzi2009,Iovino2010,Koyama2010}.  

Motivated by the rapid evolution of the spiral population since $z<1$
\citep{Dressler1997,Kodama2001,Postman2005,Wilman2008}, many investigations
focus on the transformation of infalling spirals, either via a rapid,
triggered burst of star formation or quenching of subsequent star
formation.  Simulations are sufficiently advanced that new insight can
be obtained by, e.g.\ comparing star-forming regions and gas-loss
rates to observations \citep{Tonnesen2007}.  Also, galaxy mass may
have a more dominant role in quenching star formation when clusters
are still assembling \citep{Peng2010,Muzzin2012}, i.e., when galaxies
are in a more group-like environment.  However, to connect galaxies
evolving in groups to more massive clusters requires identifying the
progenitors of local clusters, i.e., galaxy groups at higher redshift.

The supergroup SG1120-1202 (hereafter SG1120) at $z=0.37$ 
provides a unique laboratory for studying galaxies in a group
environment before the groups merge to form a more massive galaxy
cluster.  SG1120 was discovered in the Las Campanas Deep Cluster
Survey \citep{Gonzalez2001} and is composed of four X-ray
luminous galaxy groups that will merge to form a cluster comparable in
mass to Coma \citep{Gonzalez2005}.  Our extensive multi-wavelength
observations show that most of SG1120's members have early-type
morphologies \citep[$>60$\%;][]{Kautsch2008} and that most of the S0
members formed before the cluster phase \citep{Just2011}.  However,
the fraction of members that are bright 24$\mu$m sources is nearly as
high as in the field \citep[$\sim30$\%][]{Tran2009}.  Spectroscopy
with integral field units also show that several SG1120 members have
outflowing winds that may aid in quenching star formation
\citep{Freeland2011}.

In this paper, we focus on current star formation as traced by UV
emission using {\it Hubble Space Telescope} imaging with WFC3/F390W.
By combining with our existing ACS/F814W mosaic, we obtain high-resolution 
color maps of the supergroup members to pinpoint 
(dust-free) star-forming regions and measure internal color gradients.
We compare the UV maps to stellar masses and projected group distance
to quantify the relative importance of mass versus environmental
quenching.  Throughout the paper, we use $H_0$ = 70 km s$^{-1}$ Mpc$^{-1}$,
$\Omega_M=0.3 \text{ and }\Omega_{\Lambda} = 0.7$. At z = 0.37, this
corresponds to a scale of 5.12 kpc arcsec$^{-1}$ and a look-back time of 4
Gyr.

\section{Observations}

\citet{Gonzalez2005} identified a merging system of four X-ray
luminous groups with a total combined mass of $5.3 \times 10^{14}
M{_\odot}$ (hereafter called SG1120; see Table \ref{table:sg1120} for SG1120 properties).
The individual galaxy groups lie
on the local $\sigma - T $ relation, and a virial analysis using their
X-ray masses and relative locations indicates that these groups are
bound to each other, likely infalling for the first time, and will
merge into a single system by $z=0$. A weak-lensing analysis based on
HST imaging supports this picture \citep{Smit2015}, and 174 group
galaxies have been spectroscopically confirmed using Magellan, MMT, and
VLT/VIMOS.

Previous studies comparing SG1120 to clusters have found comparable
fractions of early-type members \citep{Kautsch2008} but also enhanced
fractions of \mipsmu\ sources \citep{Tran2009}. The high fraction of
early-type galaxies combined with enhanced star formation in SG1120
indicates that the transition to the cluster environment is well
underway even in the group environment. The combination of cluster and
field-like properties make SG1120 a unique laboratory for
investigating how environmental processes drive and/or quench star
formation.

\begin{deluxetable*}{ l c c c c c c c c c c}
\tablecolumns{8}
\tablecaption{SG1120-1202 Constituent group properties.}
\startdata
Index&	$\alpha$&		$\delta$&		$z$&	$T$&	$M$&	$\sigma$&	$R_{200}$  & $N$\\
	&	(J2000) & (J2000) &	& (keV) & $(M_\odot)$ & (km $s^{-1}$) & (kpc) & (members) \\
\hline 
\hline
1&		11:20:07.48&  -12:05:09.1& 	0.3522& 	2.2	&	1.3 $\times ~10^{14}$&	303&	240& 	28 \\
2&		11:20:13.33& -11:58:50.6&	0.3707&	1.7	&	8.0 $\times~10^{13}$&	406&	320&	27 \\
3&		 11:20:22.19& -12:01:46.1&	0.3712&	1.8	&	8.9 $\times ~ 10^{13}$&	580&	460& 	54 \\
4&		11:20:10.14&  -12:08:51.6&	0.3694&	3.0	&	2.3 $\times ~ 10^{14}$&	576&	460& 	47 
\enddata
\tablecomments{Properties summarized from \citet{Gonzalez2005} and \citet{Tran2009}}
\label{table:sg1120}
\end{deluxetable*}

\subsection{Hubble Space Telescope Imaging}

\begin{figure}[h]
\centering
\includegraphics[width=0.4\textwidth]{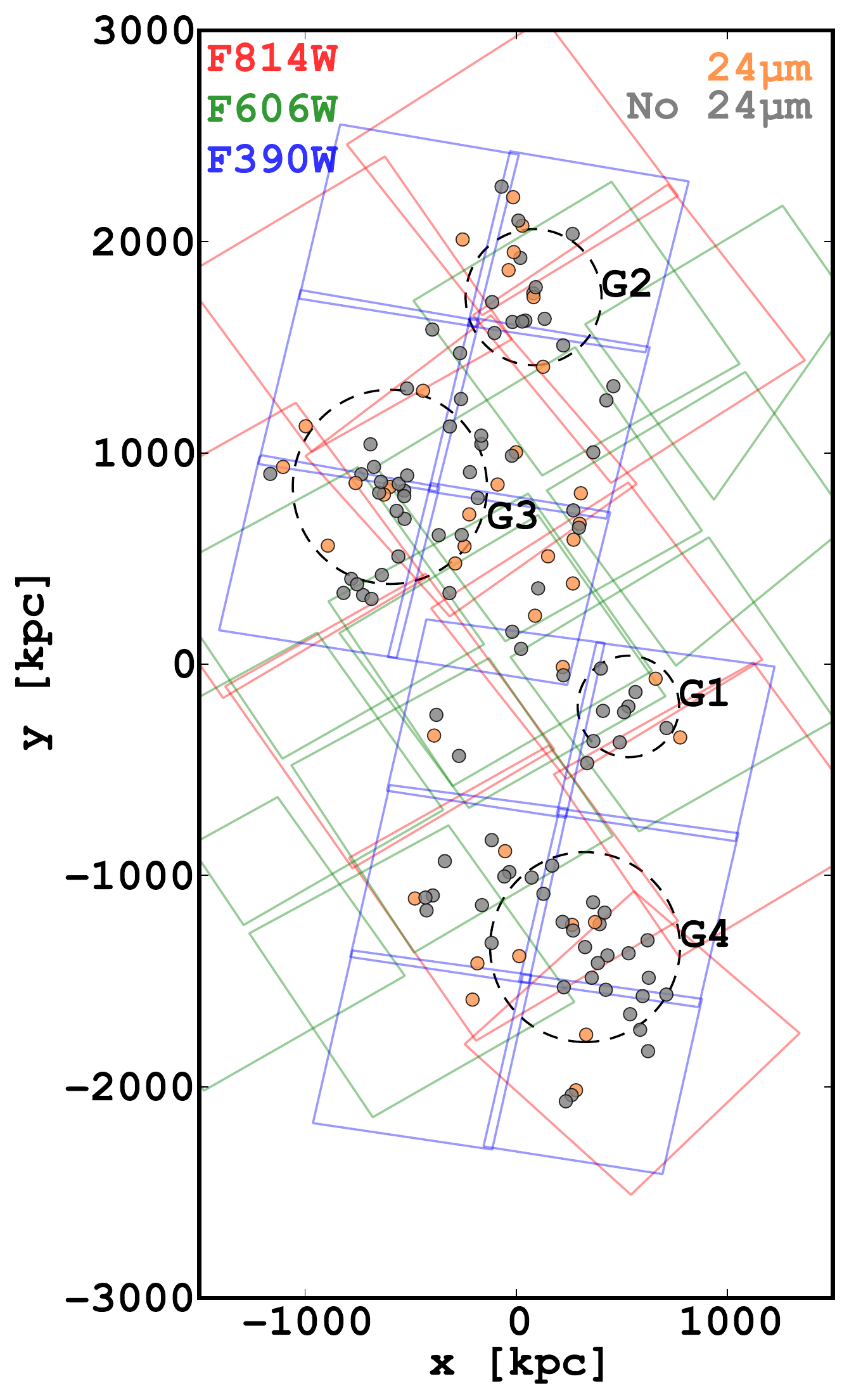}
\centering
\caption{Spatial distribution of spectroscopically confirmed
  members across the entire galaxy supergroup SG1120-1202; north is up
  and east is to the left. HST coverage in WFC3/F390W, ACS/F606W, and
  ACS/F814W are included in blue, green, and red, respectively.  MIPS
  \mipsmu\ detection or non-detection is denoted with orange or gray
  color, respectively.  Black dashed circles of radius $R_{200}$ are
  centered on each Brightest Group Galaxy (BGG), which are labeled
  according to the ordering of \citet{Tran2009}.}
\label{fig:xy_footprint}
\end{figure}

We employ HST imaging of an $\sim8'\times12'$ mosaic across three
filters: F390W (WFC3/UVIS), F606W (ACS/WFC), and F814W (ACS/WFC) for a
total of 44 pointings (combined primary and parallels) during cycles
14 (GO 10499) and 19 (GO 12470).  
The exposure times in F814W and F390W were 2000 s and 2610 s, respectively.
At $z=0.37$, these three filters
probe rest-frame UV and optical emission to track both recent star
formation and the existing stellar population. Furthermore, the high
resolution of HST/WFC3 resolves the internal structure and the fine
details of the individual galaxies.  We measure galaxy sizes (radii)
using F814W and use F390W to identify compact star-forming regions.
The F606W is only used to generate the color images for a subset of
members; the F606W coverage is not as extensive as F390W and F814W
(see Fig.~\ref{fig:xy_footprint}).

Of the original sample of 174 spectroscopically confirmed members, we
exclude 31 because they do not have imaging in either F814W or F390W.
Furthermore, eight galaxies fall near boundaries on the CCD and so our
measurements are unreliably noisy, leaving a total of 136 confirmed
supergroup galaxies for which we can measure galaxy colors and
visually classify their F390W emission.  For details on the
spectroscopy and stellar mass measurements from multi-band
ground-based observations, we refer the reader to \citet{Tran2009}.

\subsection{Measuring Galaxy Colors}  

To measure colors and fluxes via Source Extractor \citep{Bertin1996},
we first redrizzle our data to align all images.  We utilize the
default HST pipeline through the calibration and flat-fielding step
(flt output images), but redrizzle our data via AstroDrizzle
\citep{Gonzaga2012} to combine our mosaic with optimized pixel
resampling and astrometry.  
To retain resolution and to ensure detection of small-scale star forming regions, 
we apply the finest pixel resolution,
0$\farcs$0396/pixel (WFC3/UVIS), and resample the other two datasets (ACS
F606W and F814W) from their native scale (0$\farcs$05'/pixel) to match.
Resampling all images to a common resolution facilitates  
consistent color measurements across multi-band imaging. 
This resampling does not alter colors by more than 2\%. 
We use the \textit{final-wht-type} parameter to ensure our output weight
maps are inverse variance maps as per the expected input to SExtractor
 \setcitestyle{notesep={ }} 
\citep[see][for more details]{Skelton2014a}.

We measure colors with SExtractor \citep{Bertin1996} in dual-image
mode, which uses a reference image (here F814W) for detection and then
measures fluxes in the target image (F390W or F814W).  Only by using
the same detection map can we directly compare colors and sizes across
the filters.  All pixels that are associated with a galaxy in the
detection image (segmentation map) are used in calculating its flux in
the target image.

We utilize published values for instrument and filter-specific fields
such as zeropoints, FWHM, etc., and use default parameters for all
fields except minimum detection. We adjust minimum detection area to
prevent faulty detections from entering our catalogs, especially in
F390W, where object sizes are substantially smaller and the background
noisier. Our choice for detection area includes all
galaxy members, while filtering most of the false detections such as
those along the imaging boundaries and in inter-chip regions.  Table
\ref{table:results} includes the magnitudes in F814W and F390W and
half-light radius in F814W for all spectroscopically confirmed group
galaxies.

\subsection{Identifying UV Emission from Star Formation}

We visually inspect the HST imaging to identify SG1120 members that
have bright pockets of UV emission indicative of ongoing (unobscured)
star formation. The O and B type stars illuminating these regions burn
for $\sim30$ Myrs and, while they constitute less than 7\% of stellar
mass \citep{Wuyts2012}, these massive stars effectively trace new
sites of unobscured star formation.  We note that the absence of
OB-type stars does not imply that a galaxy is devoid of star
formation, e.g.\ UV light is easily reprocessed to longer wavelengths
by dust that is then detected at \mipsmu.

\begin{figure*}[h]
\centering
\includegraphics[width=\textwidth]{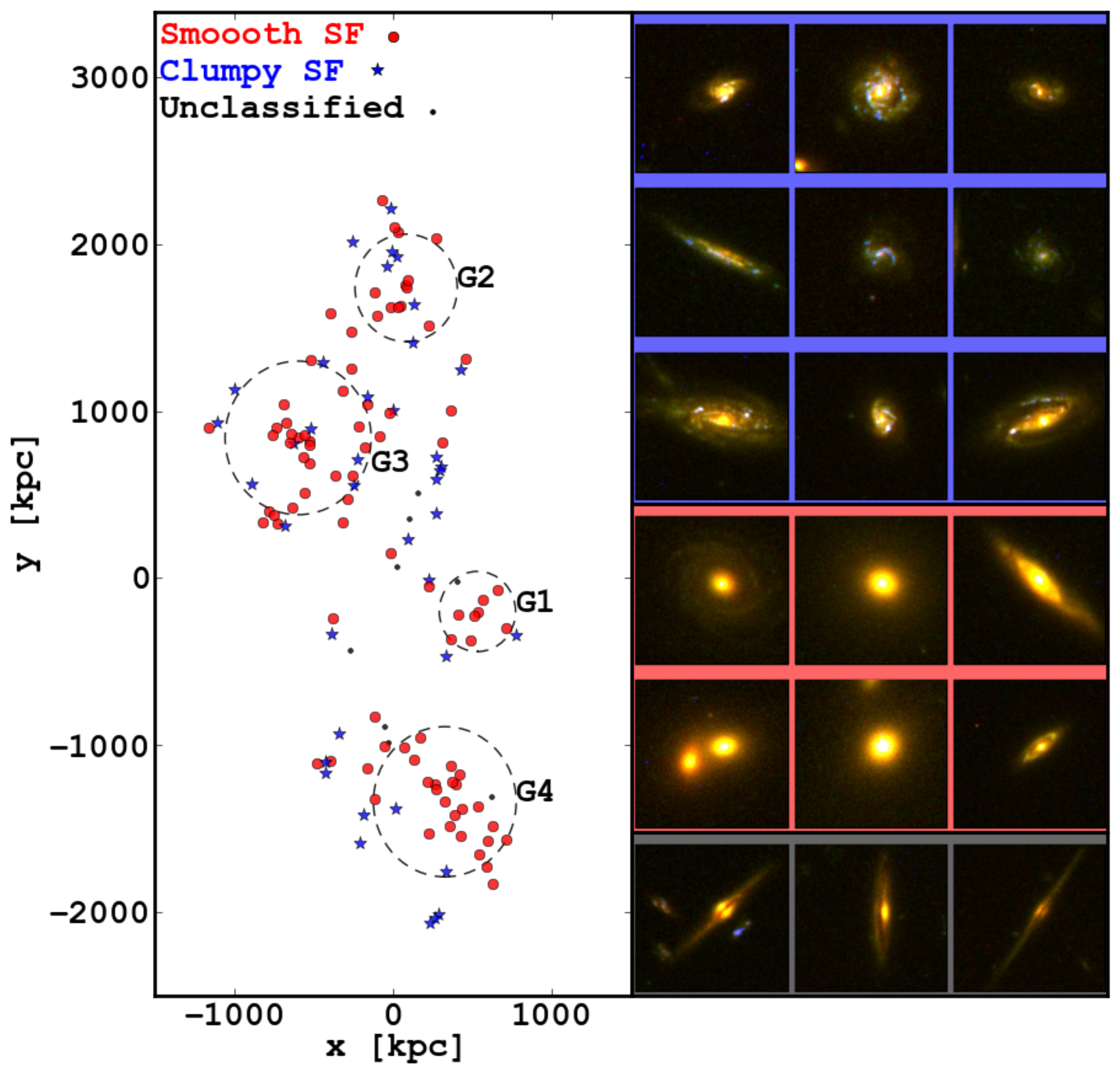}
\centering
\caption{Spatial distribution of supergroup galaxies where the
  members are separated by visual identification of UV emission as
  traced by F390W.  The dashed circles correspond to $R_{200}$ for
  each group (see Table~1). Galaxies with pockets of UV emission in
  their disks are considered ``clumpy'' and those with smooth radial
  profiles are ``smooth''.  Examples of clumpy and smooth
  classifications are shown as color images (right subpanels) that are
  generated by combining HST F814W, F606W, and F390W.  Unclassified
  objects are typically dust-obscured edge-on galaxies such as those
  in the bottom thumbnails. }
\label{fig:xy_clumpy}
\end{figure*}

J.M. and K.T. separately examined the HST imaging in the individual
filters as well as in the combined RGB images 
(see Fig.~\ref{fig:xy_clumpy}, right).   Group members are classified as
``clumpy'' if UV-bright regions exist throughout the galaxy and
``smooth'' if there are no such regions. Several galaxies are edge-on
disks and therefore cannot be reliably classified; these are denoted
as ``none.''

The right side of Figure \ref{fig:xy_clumpy} shows RGB
images for a sample of each visual classification, and Table
\ref{table:results} includes the classification for each
galaxy, including ``jellyfish'' candidates. Because of their 
 increased rest-frame UV flux, regions of ongoing (unobscured) 
 star formation are visible as blue pockets of
light (F390W) in the RGB images.  Approximately 30\% of the supergroup
have ``clumpy'' UV emission.

\begin{figure*}
\centering
\includegraphics[width=0.47\textwidth]{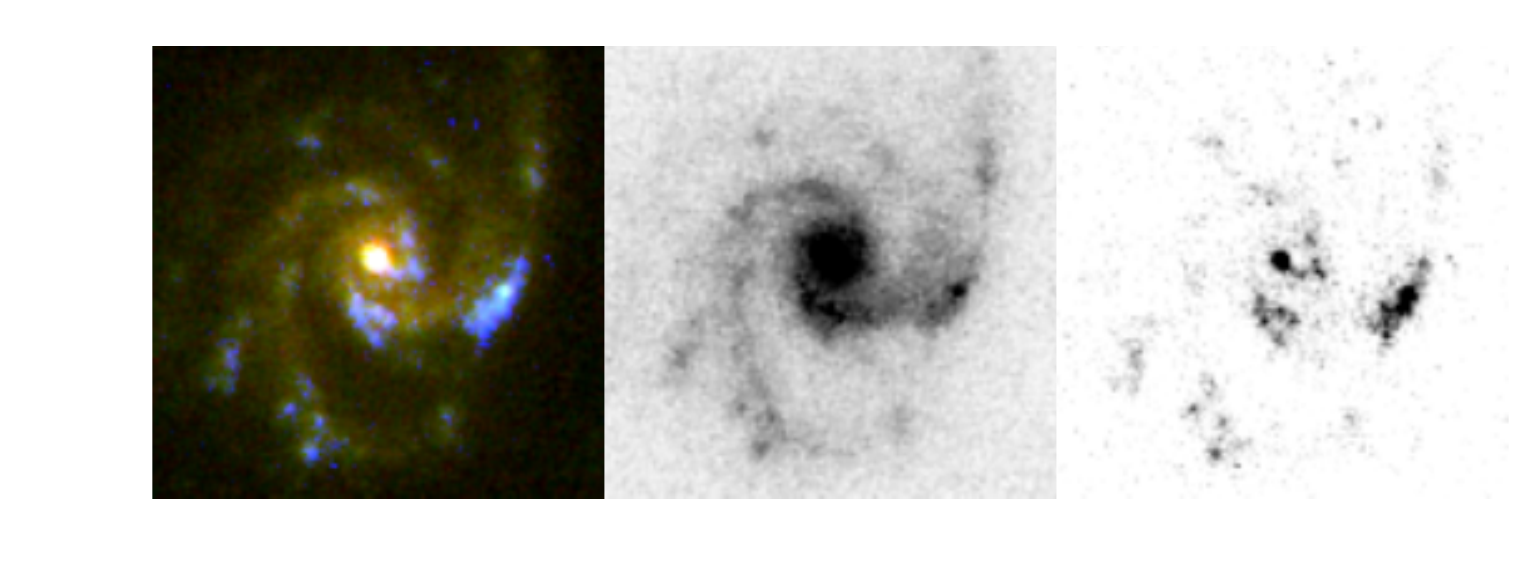} 
\includegraphics[width=0.47\textwidth]{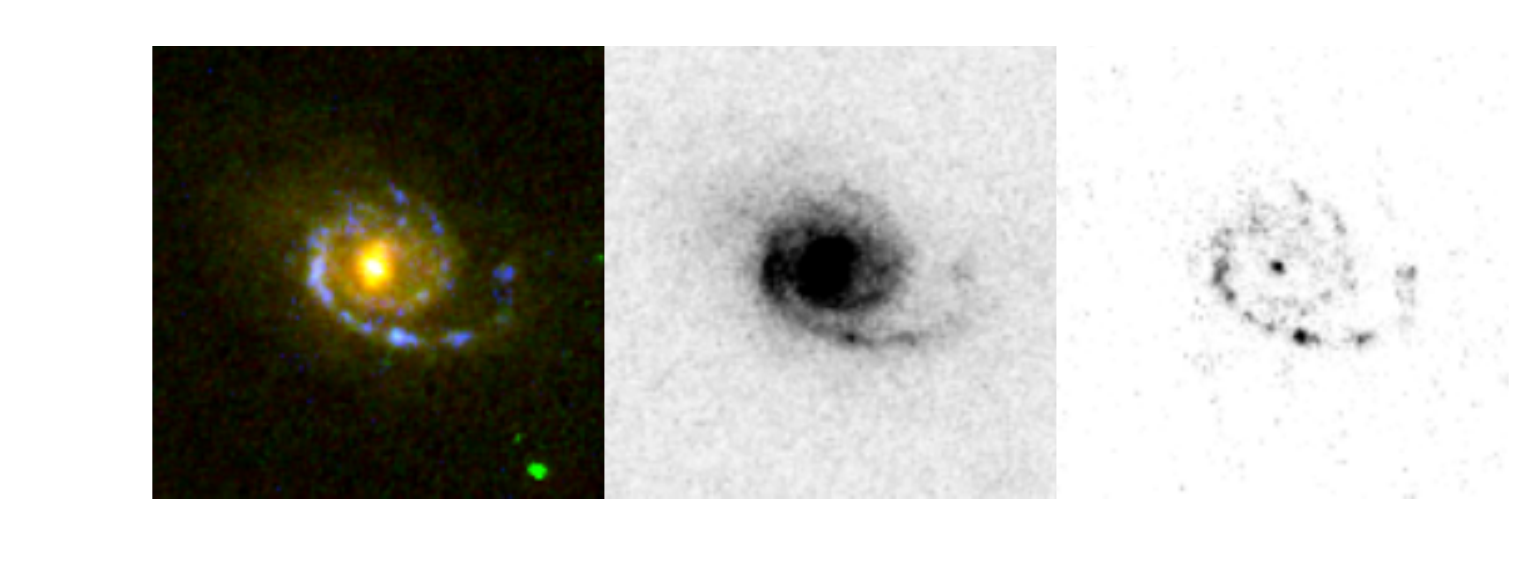} 
\caption{Examples of member galaxies (left: J112013.2-120345.2, right:J112014.5-115808.9), which we visually characterize as jellyfish candidates \citep{Smith2010}. 
Color images (left) are generated by combining HST F814W, F606W, and F390W. 
Grayscale images of F814W imaging (middle) and F390W imaging (right).
The presence of asymmetric clumps and trails of knots and filaments
serves as the defining characteristics of such galaxies. Table \ref{table:results} 
 includes labels denoting all jellyfish candidates.}
\label{fig:jellyfish}
\end{figure*}

We use the F390W imaging to visually identify jellyfish candidates,  
which are characterized by asymmetric ultraviolet clumps 
 with trails of knots and filaments \citep{Smith2010}. 
 The high resolution of HST imaging allows us to
  visually identify such features according to the criteria used 
  in the literature \citep[e.g.\ ][]{Smith2010, Owers2012, Ebeling2014}
even at $z\sim 0.37$. In Fig.~\ref{fig:jellyfish} we include
 two such examples with RGB thumbnails and grayscale images of F814W and F390W.
 The side-by-side comparison allows the rest-frame UV emission of trails 
 to directly contrast the underlying stellar population. 


\subsection{Spitzer/MIPS \mipsmu\ Fluxes}

We use the \textit{Spitzer} MIPS \mipsmu\ fluxes from
\citet{Saintonge2008} and \citet{Tran2009}.  To summarize, the
\mipsmu\ observations were retrieved from the \textit{Spitzer} archive
and spectral energy distributions were fit to determine the IR
luminosities and corresponding star-formation rates of each source 
\citep[see][for
  details]{Saintonge2008}.  At $z\sim0.37$, the 80\% completeness
limit corresponds to $\log(\mathrm{LIR})[\mathrm{ergs^{-1}}] = 43.8$
or, equivalently, a star-formation rate of $3 M_{\odot}$.  Figure
\ref{fig:xy_footprint} shows the spatial distribution of the
\mipsmu-detected group galaxies that also have the required F390W and
F814W imaging.  Following \citet{Tran2009}, we attribute the IR
emission to dusty star formation (contamination by active galactic
nuclei is $<3$\%).

\section{Results}

\subsection{UV Morphology Correlates with Projected Distance}\label{sec:uv}

\begin{figure}[h]
\centering
\includegraphics[width=0.4\textwidth]{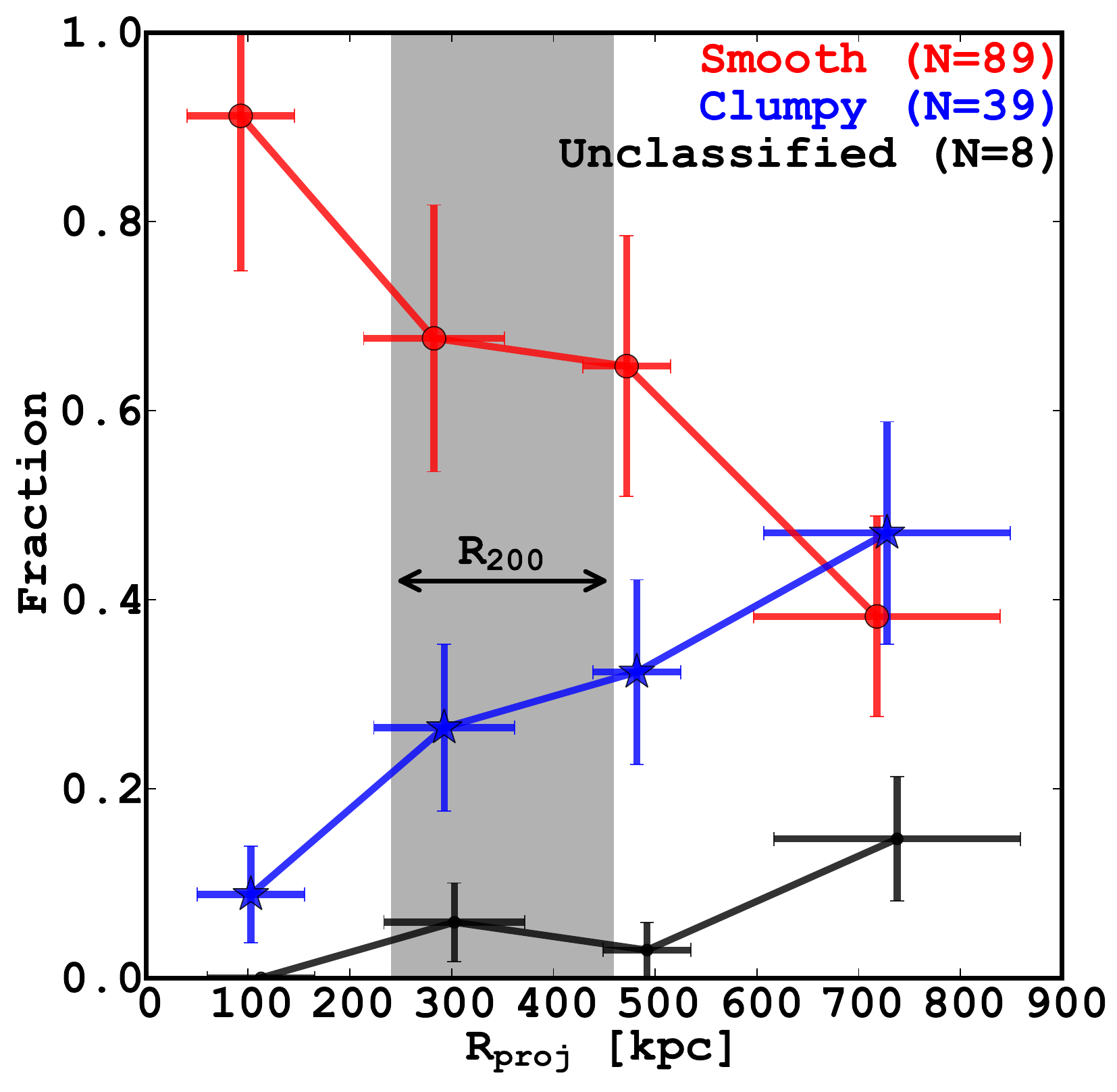}
\caption{We stack the four groups in SG1120 and show the fraction of
  members with pockets of UV emission (clumpy) relative to those
  without (smooth) as a function of the projected group-centric
  distance.  The binned points are offset by 10 kpc for clarity, and
  the $R_{200}$ range for the four groups (Table~1) is shown as a
  vertical band.  Clumpy galaxies are characterized by UV emission
  from ongoing star formation at extended radii and their fraction
  increases at larger distances from the group core.  Unclassified
  galaxies are mostly comprised of edge-on disks, where star formation
  is obscured by the disk; their fraction also increases with
  increasing distance from the core.}
\label{fig:clumpyFrac}
\end{figure}

As shown by \citet{Kautsch2008}, the trend of increasing early-type
fraction in denser environments is already established in the SG1120.
We build on this morphology--density relation by examining the spatial
distribution of supergroup galaxies with pockets of UV emission
(``clumpy''; Fig.~\ref{fig:xy_clumpy}, right).  We stack all four
galaxy groups and find that the fraction of members with patches of UV
emission (clumpy) increases with increasing distance from the group
centers (Figure~\ref{fig:clumpyFrac}).  The tendency for clumpy
members to be at larger projected distances from the cores suggests
that newly accreted group galaxies with UV-bright star-forming regions
are extinguished as they fall into the groups.  While there are a
handful of members with ``clumpy'' emission in the group cores, their
position may be due primarily to projection effects.

\subsection{Color--Magnitude Diagrams (CMDs)}

\begin{figure}[h!]
\centering
\includegraphics[width=0.4\textwidth] {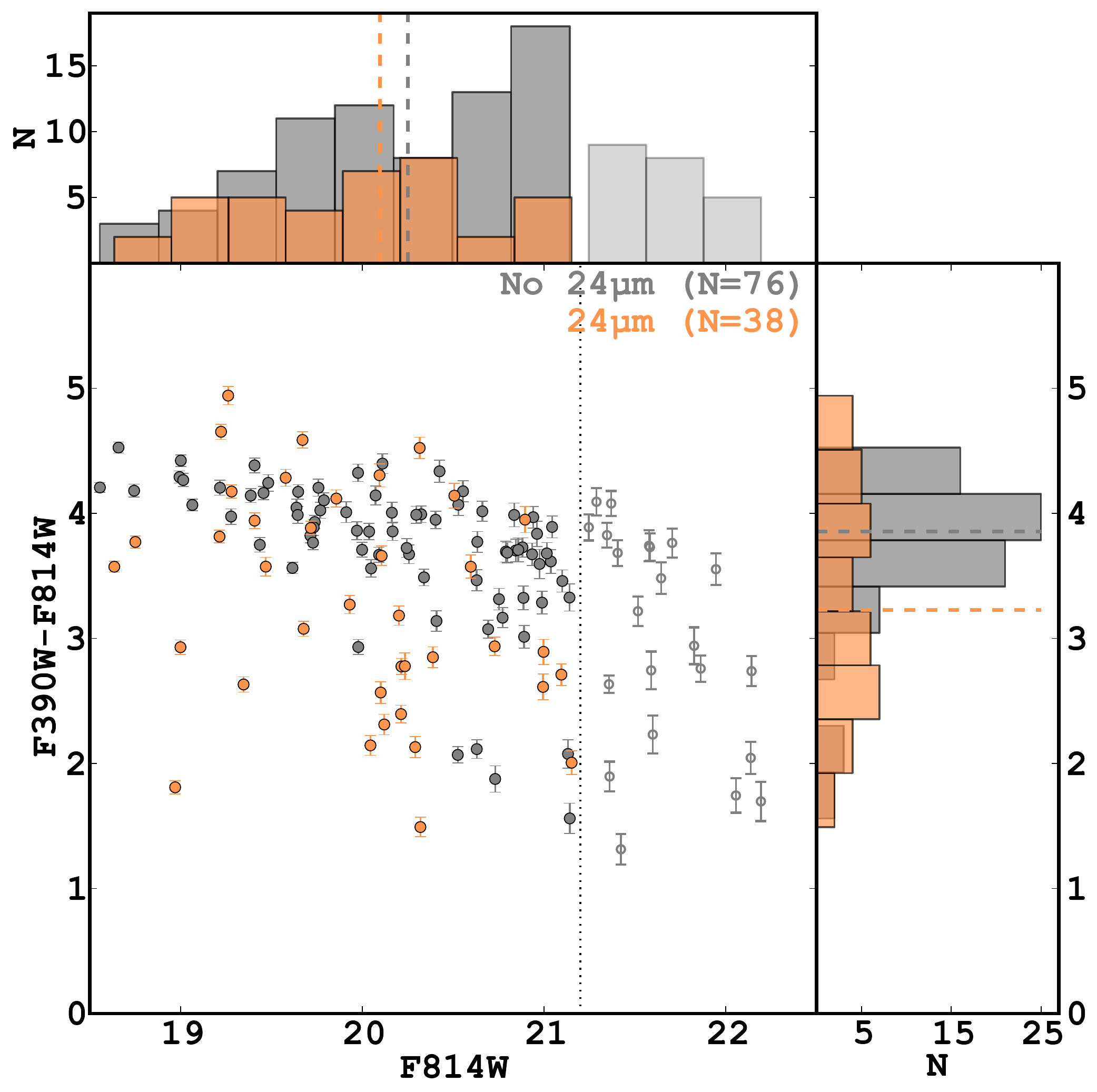}
\includegraphics[width=0.4\textwidth] {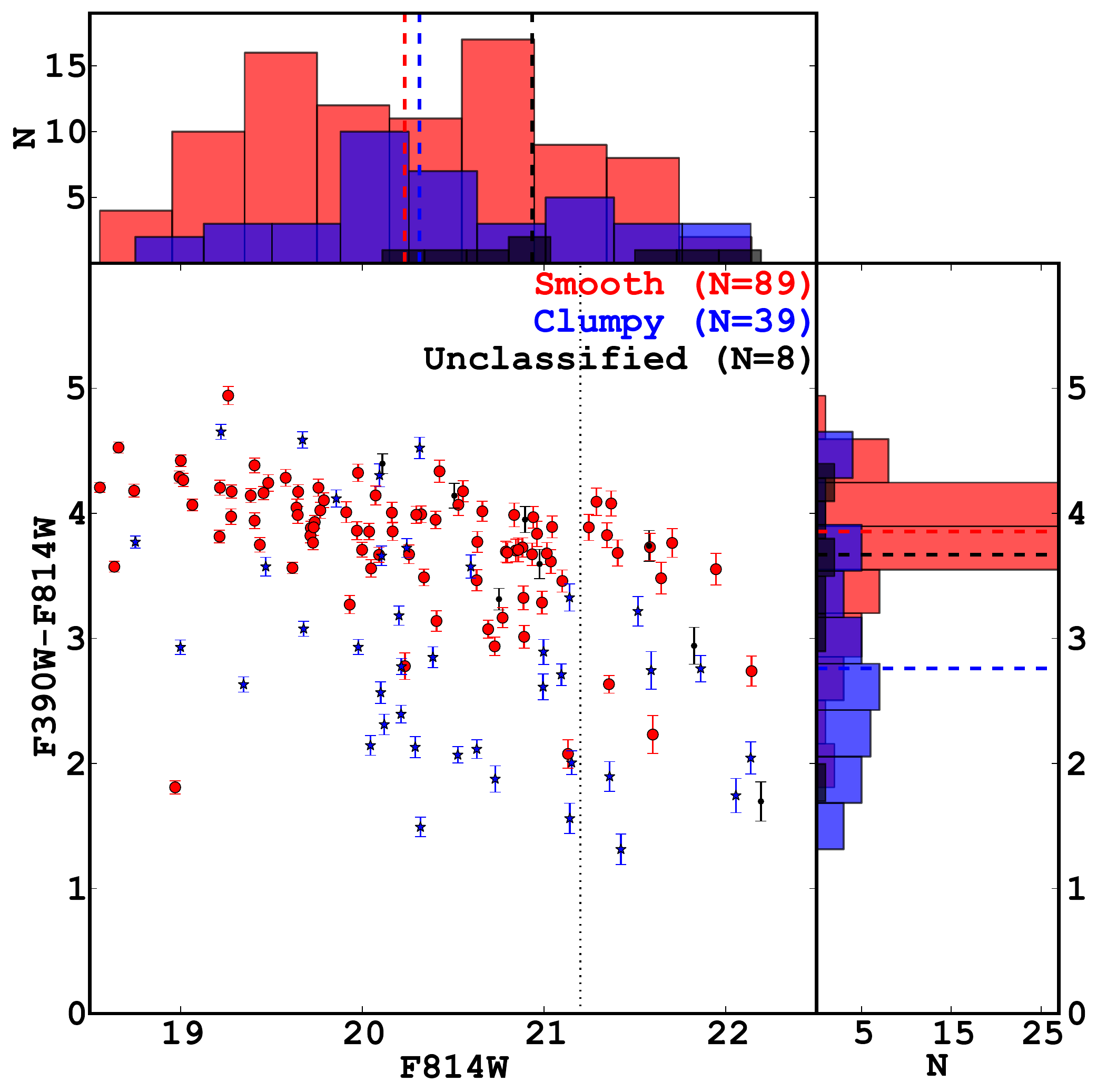}
\caption{Color--magnitude diagram (top) for supergroup galaxies
  comparing $24\mu m$ (orange filled circles) and non-detected (gray
  filled circles) members; galaxies fainter than the spectroscopic
  completeness limit of $m_{F814W} =21.2$ (dotted vertical line) are
  shown as open gray circles.  Also shown is the CMD (bottom) for
  members with pockets of UV emission (clumpy; blue stars) or smooth
  profiles (red filled circles).  The upper and right side histograms
  show the $m_{814}$ and \uvcol\ distributions for the different
  galaxy populations; the vertical dashed lines correspond to the
  average values.  There is a larger difference in average
  \uvcol\ color between clumpy vs. smooth galaxies ($\sim$1.0 dex)
  compared to $24\mu m$ vs. non-$24\mu m$ ($\sim$0.3 dex). }
\label{fig:cmd}
\end{figure}

CMDs are an efficient method for tracing how
a galaxy evolves as its stellar population ages.  As a galaxy's star-formation rate decreases, its colors redden and move it from the blue
cloud to the red sequence within the CMD.  Rest-frame unobscured UV
emission is particularly effective at measuring unobscured SF rates
because it is generated by the youngest and most massive main-sequence
OB-type stars with lifetimes of $<30$~Myr.  In Figure \ref{fig:cmd},
we compare the \uvcol\ distributions for different populations of
supergroup members based on their IR (\mipsmu) and UV (F390W)
emission.

In Fig.~\ref{fig:cmd} (top), we compare the color distribution of
\mipsmu\ members to those that are undetected.  Although \mipsmu\ and
UV flux both trace star formation \citep{kennicutt2012}, the
correlation with \uvcol\ colors can be weak because UV emission is
easily reprocessed by dust, which then emits in the far-infrared
\citep{meurer1999}.  The 28 IR-detected supergroup galaxies above our
completeness limit of $m_{F814W} > 21.2$ are about 0.3 dex bluer
than the non-IR members, and a KS test confirms that the difference is
significant ($>3\sigma$).  Note that while the \mipsmu\ members tend
to be blue, they also span the full range in \uvcol\ color and include
some of the reddest galaxies.

The transition from star-forming to quiescent galaxy also usually
corresponds to changes in morphology from late- to early-type
\citep{vanDokkum1998a,Balogh2004b,Bell2004}.  We compare the 
supergroup galaxies based on their clumpy/smooth classification
(Figure \ref{fig:cmd}, bottom) and find that these two populations
differ in average \uvcol\ color by $\sim1$~dex, i.e.\ more than three 
times larger than the IR vs. non-IR members.  When considering all of
the members, the average \uvcol\ colors for smooth and clumpy are
$3.76\pm0.08$ and $2.79\pm0.09$ respectively; if we consider only
$m_{F814W}<21.2$, the average \uvcol\ are $3.81\pm0.08$ and
$2.92\pm0.09$.

To summarize, the supergroup galaxies with smooth light profiles tend
to be very uniform in color, to have color shallow gradients, to be red, to have early-type morphologies,  and to populate the group cores \citep[Figs. \ref{fig:xy_clumpy} \&	
\ref{fig:clumpyFrac},][]{Kautsch2008}.  In contrast, the clumpy galaxies contain both
red and blue components, span the range in galaxy color, tend to have
prominant disks, and are mostly outside the group cores.

\subsection{Total Galaxy Color Correlates with Projected Distance}\label{sec:color_rproj}

\begin{figure}[h]
\centering
\includegraphics[width=.4\textwidth]{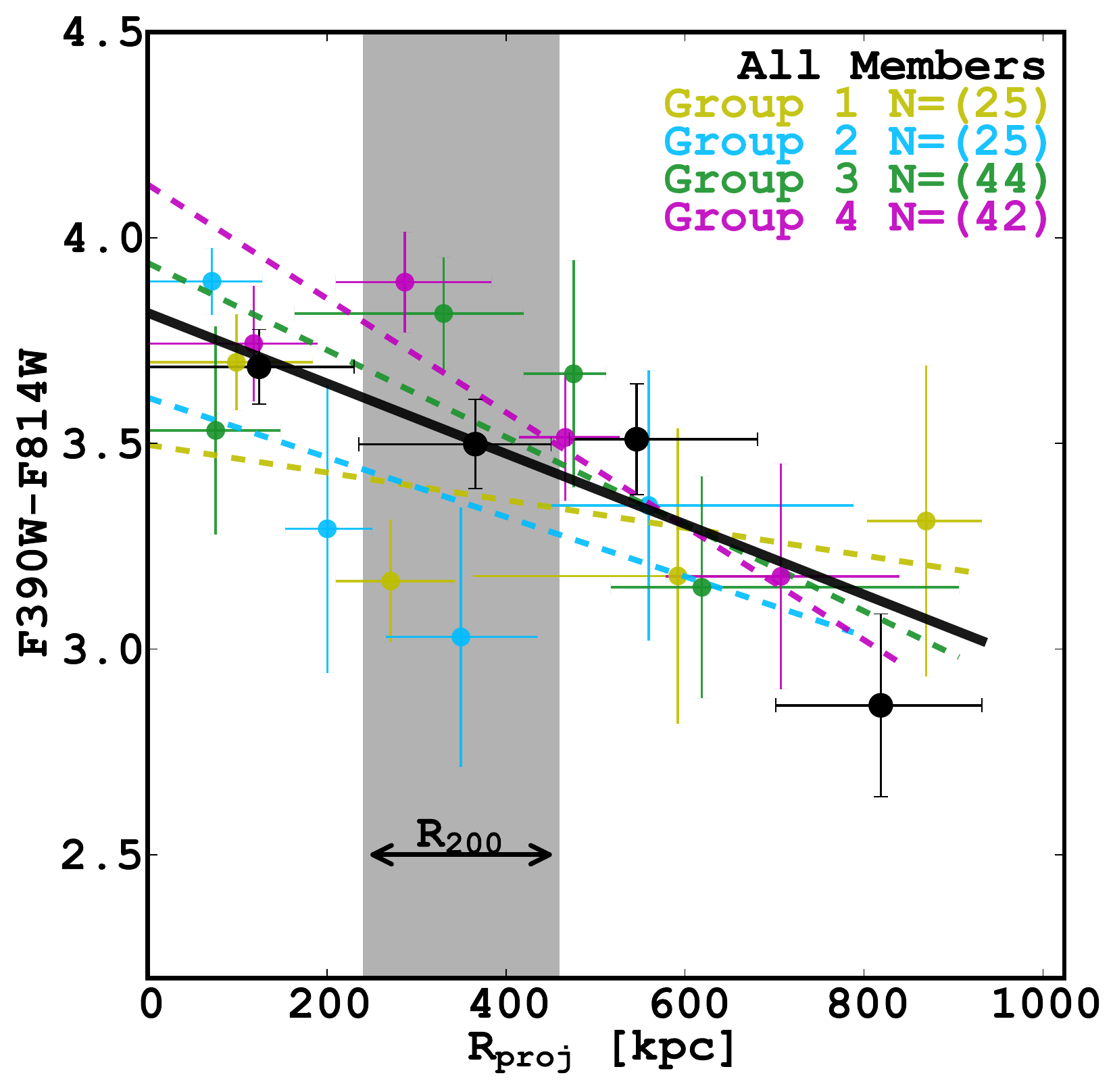}
\caption{Total \uvcol\ color for individual supergroup galaxies
compared to their projected distance (Rproj) for the group stack; the
vertical dashed line corresponds to a projected distance of 400 kpc
which is approximately $R_{200}$ (see Table~1).  We measure a strong
correlation ($>3\sigma$ confidence) between \uvcol\ color and
projected distance (\Rproj): members are redder in the group cores.
The error bars in \Rproj\ and \uvcol\ color represent, respectively,
the bin range and standard deviation within the bin.
\label{fig:ui_rproj} }
\end{figure}

In Fig.~\ref{fig:ui_rproj}, we compare how total \uvcol\ color and
local environment are correlated by stacking the four groups into a
single system and using the projected radius (\Rproj) from the 
nearest Brightest Group Galaxy (defined as the group center). 
 The projected radius serves as a proxy for environmental density.
  For an analysis of how the \mipsmu\ emission depends on local environment 
  in SG1120, we refer to \citet{Tran2009}.

Supergroup galaxies in the group cores are redder than those at larger
distances.  A least-squares fit to the total galaxy color and
projected distance for the stacked groups (Fig.~\ref{fig:ui_rproj},
black line) measures a strong correlation of
\begin{equation}
\mathrm{(F390W - F814W)} 
= 8.56 \times10^{-4} \left(\frac{R_{\mathrm{proj}}}{ \mathrm{kpc}}\right) + 3.82 
\end{equation}
at $>3\sigma$ confidence (errors in slope and offset are 2.64 and 0.12, respectively).  Least-squares fits to the individual groups
measure the same general trend.  Figure~\ref{fig:ui_rproj} includes
the error bars that represent each bin's range in \Rproj\ and standard
deviation in \uvcol.  As galaxies approach their group cores, their
increasingly red \uvcol\ colors indicate that their star formation is
quenching.

\subsection{Correlations with Stellar Mass}

\begin{figure}[h]
\centering
\includegraphics[width=.4\textwidth]{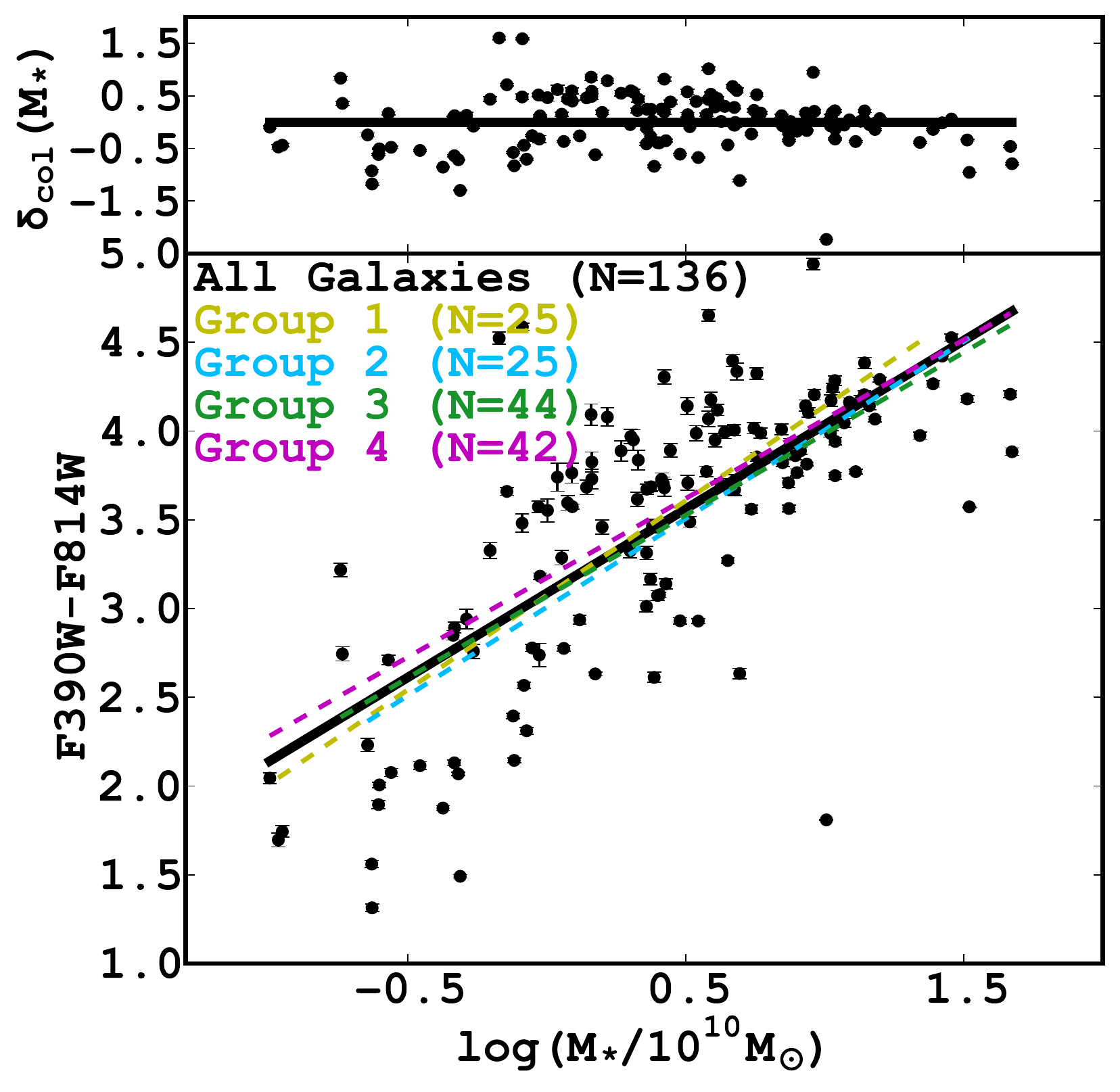}
\caption{In SG1120, the total \uvcol\ color for individual members is
  tightly correlated with their stellar mass in all of the groups
  (bottom; colored lines are least-squares fits to the separate
  groups).  A least-squares fit to the supergroup stack (black line)
  has mean residuals within $\sim0.5$~dex (top).  The strong
  correlation between \uvcol\ color and stellar mass is what drives
  the observed correlation between \uvcol\ and projected group
  distance (Fig.~\ref{fig:ui_rproj}): massive supergroup 
  galaxies tend to be red and tend to be in the group cores.}
\label{fig:mass_color} 
\end{figure}

To test if the correlation between total galaxy color and projected
distance in SG1120 (see \S\ref{sec:color_rproj}) is driven by an
increasing number of massive, passive galaxies in the group cores, we
compare \uvcol\ color to stellar mass (\Mstar) in
Fig.~\ref{fig:mass_color}.  A least-squares fit confirms a strong
correlation of

\begin{equation}
\mathrm{(F390W-F814W)} = 0.95 \times \log[M_{\star}/10^{10}M_{\odot}]+3.09
\end{equation}
(errors in slope and offset are 0.08 and 0.06, respectively) shown as a 
solid line in Fig.~\ref{fig:mass_color}: more massive
supergroup members have redder \uvcol\ colors.
Although 22 galaxies are below our spectroscopic completeness limit (which we take to estimate our mass completeness), the slope of Eq.~2 changes by less than $1 \%$ when we exclude galaxies below this limit.

\begin{figure}[h]
\centering
\includegraphics[width=0.4\textwidth]{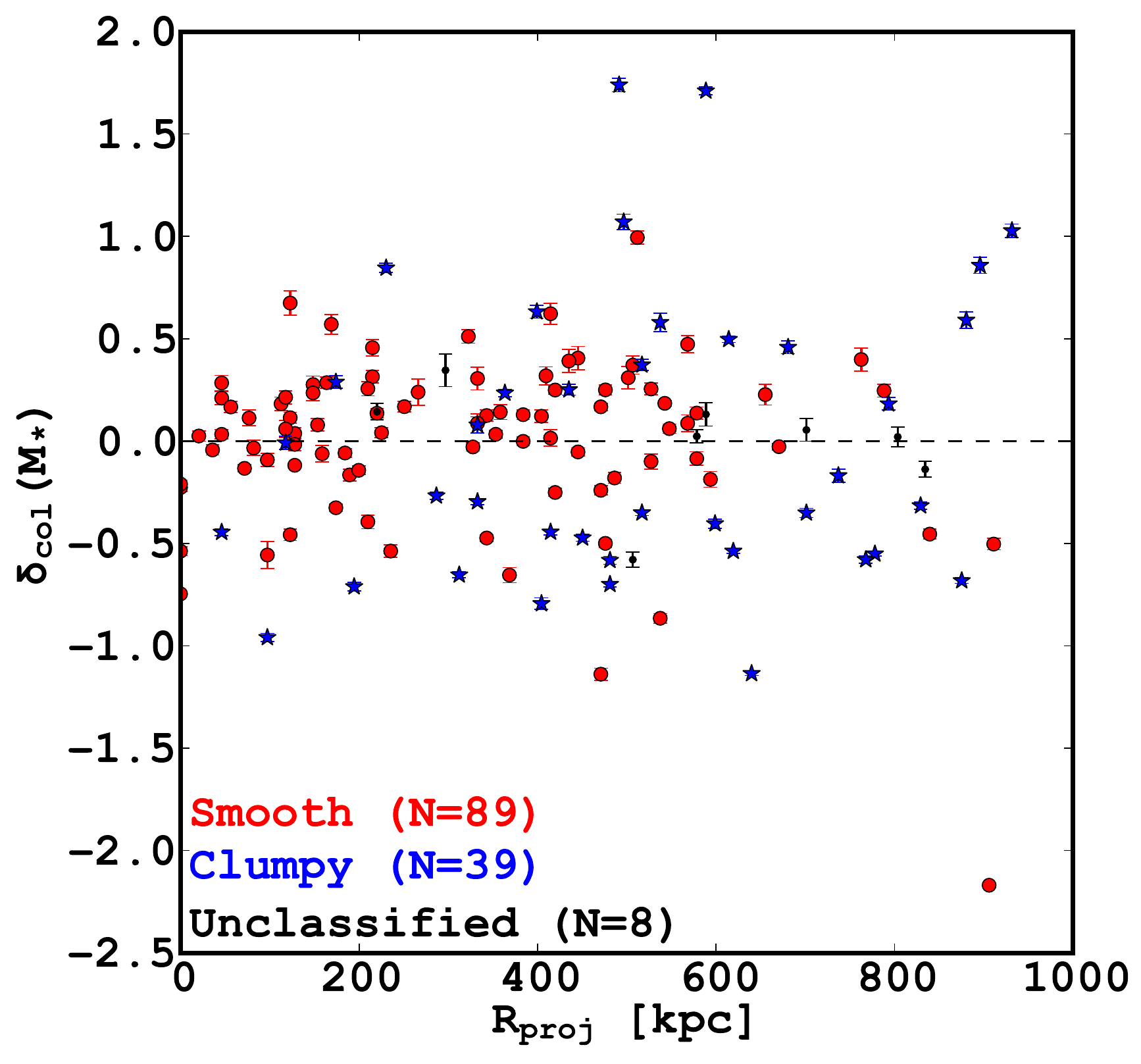}
\caption{Color deviation \duvcol\ is the difference between a
  galaxy's measured \uvcol\ color and the expected value based on its
  stellar mass (Eq.~2).  Correcting for the stellar mass dependence
  removes any trend betwen \duvcol\ and \Rproj; this also holds for
  the clumpy (stars) and smooth (large circles) members. Least-squares
  fits to \duvcol--\Rproj\ for the different galaxy populations
  confirm slopes within $1\sigma$ of zero (slope of $-2.36 \times 10^{-5} \pm 1.84 \times 10^{-4}$). 
 Thus the correlation  between total galaxy color and stellar mass explains the correlation
  between total galaxy color and projected group distance
  (Fig.~\ref{fig:ui_rproj}). }
\label{fig:residual_rproj} 
\end{figure}

We now combine the measured relation between total galaxy color and
stellar mass (Fig.~\ref{fig:mass_color}) with the measured relation
between total galaxy color and projected group distance
(Fig.~\ref{fig:ui_rproj}).  
Mass and projected group distance correlate with a slope 	
of $-9.02\pm1.95 \times 10^{-4}$, but we seek to separate the effects 
of these properties on total galaxy color.	
For each supergroup galaxy, we calculate a
color deviation \duvcol\ defined as the difference between the
galaxy's measured \uvcol\ color and the expected value given its
stellar mass from Eq.~2.  Once we account for the correlation between
total galaxy color and stellar mass, the trend between \uvcol\ color
and \Rproj\ disappears (Fig.~\ref{fig:residual_rproj}).  Least-squares
fits to \duvcol\ and \Rproj\ for the different galaxy populations
confirm slopes within $1\sigma$ of zero 
(all galaxies: $-2.36\times 10^{-5} \pm 1.84 \times 10^{-4}$),   
 i.e., the observed trend 
between \uvcol\ color and \Rproj\ is driven by stellar mass.

We test whether \Mstar\ can also explain the decreasing fraction of members
with pockets of UV emission (clumpy) with decreasing group distance
(Fig.~\ref{fig:clumpyFrac}; \S\ref{sec:uv}).  The clumpy members tend to
have lower stellar mass compared to the smooth members with average
$\log$(\Mstar/\Msun) of $9.9\pm0.48$ and $10.7\pm0.51$, respectively.
Least-squares fits to \duvcol\ and \Rproj\ of both clumpy and smooth
members confirm trends consistent with zero, 
(clumpy: $4.39\pm4.48 \times 10^{-4}$, smooth: $-2.33\pm2.00 \times10^{-4}$), 
i.e.\ the correlation 
between UV morphology and projected group distance disappears once we
account for stellar mass.


To confirm that stellar mass is the key parameter driving this relation, we test whether projected group-centric distance provides an equally good explanation of the observed correlation. The correlation between color and radius is much weaker than the trend with mass. If we define \duvcolRproj\ as the difference between the galaxy's measured \uvcol\ color and the expected color given its \Rproj\ from Eq.~1, then the residual scatter is appreciably larger (0.74 versus 0.52) and \duvcolRproj\ is strongly correlated with stellar mass (slope of $0.83 \pm 0.08$).

We find that the primary predictor for a galaxy's total \uvcol\ color
and UV morphology is its stellar mass, i.e.\ mass-quenching
\citep{Peng2010}.  
The increasing fraction of massive galaxies in the core is due to this environmental effect, 
whereas the increasing fraction of passive galaxies merely reflects mass.
Our results in the group environment of SG1120 mirror results by
\citet{Muzzin2012} for galaxy clusters at $z\sim1$.

\subsection{Internal Color Gradients}

With high-resolution imaging from HST, we map internal color gradients
at multiple half-light radii for individual supergroup galaxies.  We
measure the \uvcol\ at three circular annuli corresponding to 1.0, 1.75, and 2.5 half-light radii, where \rhalf\ is measured in the F814W
imaging.  While other studies use more annuli to measure internal
color gradients \citep[e.g.\ ][]{Tamura2000,Welikala2012,Allen2015},
we focus our analysis on three radii to ensure a robust comparison of
the galaxies' disks to their centers, while ensuring
 that measurement apertures are larger than the PSF for both filters.

\begin{figure}[h]    
\centering
\includegraphics[width=0.4\textwidth] {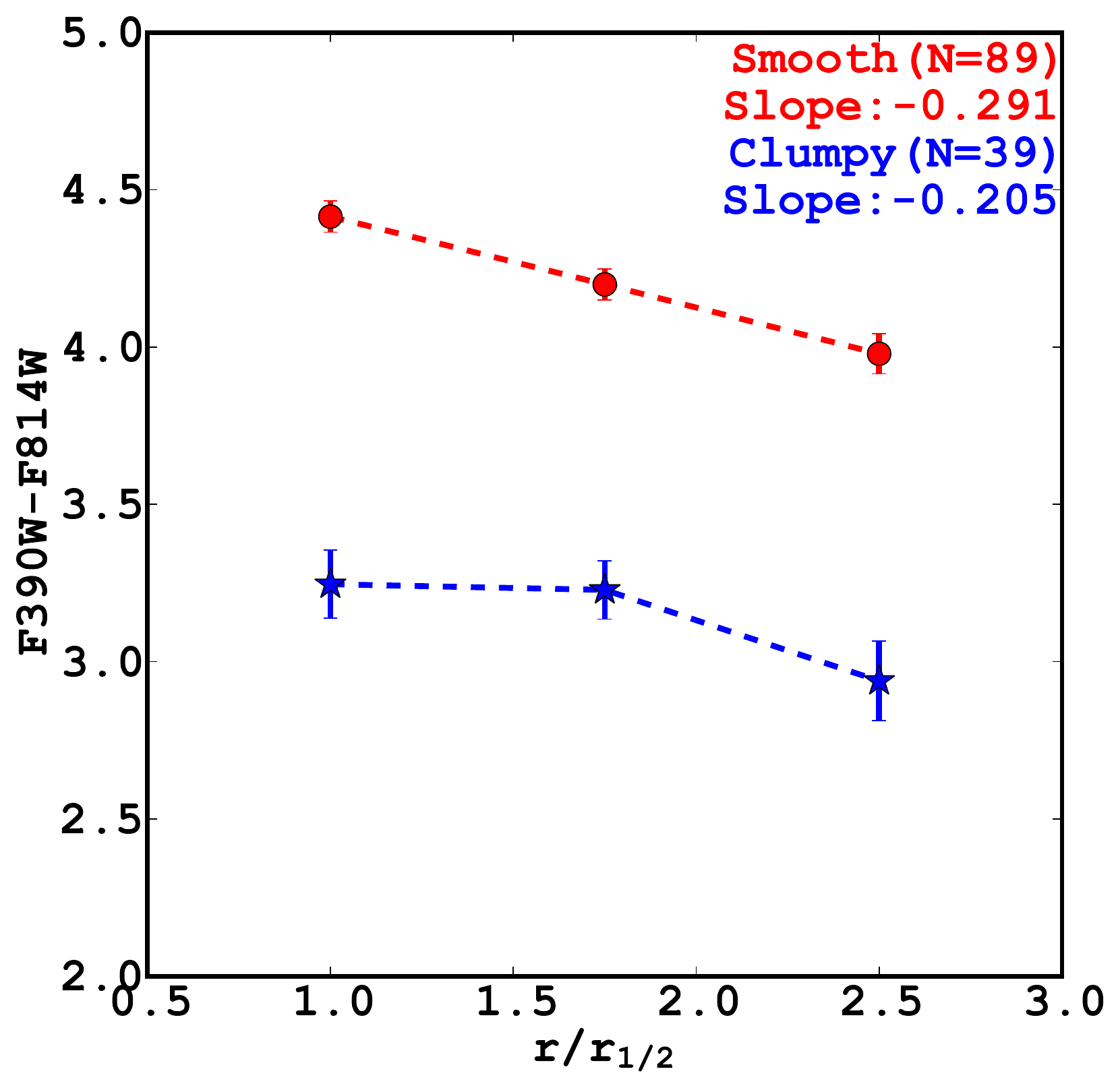}
\caption{With HST's high spatial resolution, we are able to measure
  internal color gradients and show the stacks for clumpy (stars) and
  smooth (circles) supergroup members. The smooth members
  are $\sim$1 dex redder, and both smooth and clumpy members have shallow gradients.
  }
  \label{fig:col_grad}
\end{figure}

We stack internal color gradients for the clumpy and smooth supergroup
galaxies (Fig.~\ref{fig:col_grad}) and confirm that smooth members are
$\sim 1$ dex redder than clumpy members.  The average \uvcol\ color in 
each annulus is measured from the galaxy distribution (either clumpy
or smooth) and the corresponding error is determined from 1000
boot-strapped realizations.  
The gradients of clumpy and smooth galaxies have slopes of $-0.205 \pm  0.104$ 
and $-0.281 \pm  0.002$, respectively. 

We also stack and measure internal color gradients for the
\mipsmu-detected and non-IR members.  As expected, the
\mipsmu\ members are bluer \uvcol\ throughout their galaxy disks and
centers compared to the non-IR members.

\section{Discussion}

A number of physical mechanisms continue to be explored to explain the
changes in galaxy properties with environment.  Ram pressure stripping
\citep{Gunn1972}, tidal interactions with other galaxies or with the
cluster potential \citep{Moore1998a}, morphological quenching
\citep{Martig2009}, and environmental strangulation \citep{Larson1980}
each provide different processes that can explain the observed
differences in UV emission \citep[see][for a comprehensive review of
  such mechanisms]{Boselli2006}.  However, given the breadth of these
studies and the range of (sometimes) contradictory conclusions, we do
not attempt to solve this issue definitively here with only one
system.

Instead, we consider a few clues from supergroup SG1120.  The galaxies
are spatially segregated by UV emission even before the cluster forms
(Figs.~\ref{fig:xy_clumpy} \& \ref{fig:ui_rproj}), and stellar mass is
the key.  Most of the members with pockets of UV-bright star formation
(``clumpy'') tend to be fainter (lower-mass) galaxies (Fig.~\ref{fig:cmd}), and 
a statistically significantly trend between \uvcol\ color and stellar mass (Fig.~\ref{fig:mass_color}, Eq.~2) drives this relation.
Both results reinforce the link between increasing stellar mass and decreasing star
formation \citep[``mass-quenching'';][]{Peng2010}. 

However, this is not to say that environment is unimportant
\citep[see also][]{Cooper2010}.  First, there is mass-segregation in
the group environment where the fraction of massive galaxies increases
with decreasing \Rproj.  Second, we find for the first time group
galaxies with UV tails and asymmetric UV emission
(Fig.~\ref{fig:xy_clumpy}, top 3 rows) similar to those of
jellyfish galaxies.  
These jellyfish candidates are interpreted 
as signs of shocked gas fronts in merging clusters \citep{Owers2012}
which is consistent with the incipient merger of the four X-ray
luminous galaxy groups that make up SG1120.  Third, the slope of the
UV color-stellar mass relation in SG1120 may be different from that of
the field.

The increasing fraction of UV emitting/clumpy members with increasing
projected group distance (Figs.~\ref{fig:clumpyFrac} \&
\ref{fig:ui_rproj}) mirrors the same trend for \mipsmu\ members
\citep{Tran2009}.  The shallow internal color gradients
(Fig.~\ref{fig:col_grad}) of both the clumpy and smooth populations
suggest that quenching should happen uniformly throughout a galaxy's disk and core
once it is within $\sim2\times R_{200}$ of the group core
(Fig.~\ref{fig:clumpyFrac}).  The UV emitting galaxies tend to have
lower stellar masses and can thus fade into the faint quiescent
members that are prevalent in low-redshift clusters
\citep{DeLucia2007}.  Alternatively, these quenched galaxies can merge
to form more massive (quiescent) members in the group cores
\citep{Tran2008}.


\section{Conclusions}

We map the UV emission in a supergroup at $z=0.37$ by
combining high-resolution imaging from the {\it Hubble Space
  Telescope} with extensive ground-based multi-band imaging and
spectroscopy.  SG1120 is composed of four X-ray luminous galaxy groups
that will merge and form a galaxy cluster by $z\sim0$
\citep{Gonzalez2005}.  We use WFC3/F390W and ACS/F814W imaging of
supergroup galaxies to measure total \uvcol\ colors, visually separate
UV morphology into ``clumpy'' and ``smooth'' systems, and measure
internal color gradients.  

These observations indicate that stellar mass is the primary predictor  
of UV emission in the supergroup galaxies (``mass-quenching'') while 
environment drives the higher fraction of massive galaxies in the group cores.  

We show for the first time that several group members have UV
morphologies similar to the jellyfish galaxies in massive X-ray 
luminous clusters.  The incipient merger of the four groups in SG1120
is consistent with the interpretation that jellyfish galaxies are
signs of gas shocks due to mergers.

Approximately 30\% (38/128) of the visually classified supergroup
galaxies have pockets of UV emission (``clumpy'';
Fig.~\ref{fig:xy_clumpy}), a fraction that is comparable to the
\mipsmu\ members \citep[32\%;][]{Tran2009}.  The clumpy UV members
have shallow internal \uvcol\ color gradients 
(Fig.~\ref{fig:col_grad}). 

To measure UV properties as a function of projected group distance
(\Rproj), we stack the four galaxy groups into a single system.  We
find that the UV emission, as measured by the total \uvcol\ color and
by the clumpy/smooth classification, is strongly correlated with
projected group distances (Figs.~\ref{fig:clumpyFrac} \&
\ref{fig:ui_rproj}).  We show that both of these trends in UV with
\Rproj\ are driven by the strong underlying correlation between
\uvcol\ and stellar mass: group galaxies with UV emission tend to be
lower luminosity/lower-mass members, and these systems are more common
at \Rproj$>$R$_{200}$ (Figs.~\ref{fig:mass_color} \&
\ref{fig:residual_rproj}).

However, we do find evidence of environmental processes.  Most
importantly, the higher density environment of the group cores
(\Rproj$<R_{200}$) means an increase in the fraction of members that
are massive and red/smooth (mass-segregation).  Also, the supergroup
contains jellyfish candidates whose UV morphologies likely are 
shaped by interactions with the intragroup gas.

Our analysis of the supergroup galaxies indicate that stellar mass is
the primary predictor of a galaxy's UV properties (mass-quenching),
but it is environment that regulates the galaxy mass distribution and
shapes jellyfish galaxies. 

\acknowledgements

We are grateful to J. Moustakas and D. Zaritsky for their
contributions to the SG1120 data analysis and catalogs.  We thank
L. Alcorn, B. Forrest, and Jimmy for helpful comments on the data
analysis and manuscript.  J. M.  and K. T. acknowledge support
for Program number HST-GO-12470 provided by NASA through a grant from
the Space Telescope Science Institute, which is operated by the
Association of Universities for Research in Astronomy, Incorporated,
under NASA contract NAS5-26555.

{\it Facilities:} \facility{VLT (VIMOS)}, \facility{VLT (FORS2)},
\facility{Magellan (LDSS3)}, \facility{KPNO (Mayall 4m)},
\facility{HST (ACS, WFC3)}, \facility{SST (MIPS)}, \facility{CXO
  (ACIS)}, \facility{Keck (LRIS)}.

\bibliographystyle{aasjournal.bst}
\bibliography{library}

\include{tab2}

\end{document}

%% file: tab2.tex
\LongTables
\begin{deluxetable*}{l c c c c c c c c c c c c}
\tablecolumns{10}
\tablecaption{Observed Properties of Supergroup Members  \label{table:results}}
\startdata
\hline
\hline
	Name\tablenotemark{a} &  $z$  &  $R_{\mathrm{proj}}$  &  $24\mu m$\tablenotemark{b}  &  Half-light Radius\tablenotemark{c} &  $m_{814}$\tablenotemark{d} & $\Delta m_{814}$  &  $m_{390}$\tablenotemark{d}  &  $\Delta m_{390}$ & $\log [M_{\star}/M_{\odot}]$\tablenotemark{e} &  Clumpy \\
	 & \phantom{z} & (kpc) & (mJy) & F814W (kpc) & \phantom{mag} & \phantom{magErr} & \phantom{mag} & \phantom{magErr} & \phantom{mass} & \phantom{clumpy} \\
\hline
\\
J112023.8-120326.1	&0.3708	&527	&\nodata	&3.22	&19.76	&0.002	&23.96	&0.029	&11.0	&Smooth\\
J112019.3-120516.9	&0.3545	&911	&\nodata	&1.45	&20.69	&0.002	&23.77	&0.028	&10.4	&Smooth\\
J112028.8-120127.6	&0.3403	&517	&1.96	&3.52	&19.68	&0.002	&22.75	&0.013	&10.4	&Clumpy\\
J112027.4-120050.0	&0.3527	&491	&0.24	&3.91	&20.32	&0.003	&24.84	&0.033	&9.8	&Clumpy\\
J112026.0-120240.4	&0.3733	&404	&0.12	&3.62	&20.99	&0.005	&23.61	&0.028	&10.4	&Clumpy\\
J112021.3-120149.6	&0.3677	&72	&\nodata	&3.40	&18.74	&0.001	&22.93	&0.018	&11.5	&Smooth\\
J112022.2-120146.0	&0.3713	&0	&0.13	&1.44	&19.72	&0.001	&23.60	&0.021	&11.7	&Smooth\\
J112023.4-120106.7	&0.3719	&220	&\nodata	&2.79	&19.22	&0.002	&23.42	&0.023	&11.1	&Smooth\\
J112023.9-120134.3	&0.3687	&148	&\nodata	&1.20	&21.41	&0.004	&25.09	&0.041	&10.1	&Smooth\\
J112012.2-120736.3	&0.3717	&415	&\nodata	&1.50	&21.37	&0.004	&25.45	&0.051	&10.2	&Smooth\\
J112011.5-120432.6	&0.3693	&363	&0.25	&2.90	&20.20	&0.003	&23.38	&0.018	&10.0	&Clumpy\\
J112009.7-120810.2	&0.3672	&215	&\nodata	&2.23	&19.41	&0.001	&23.79	&0.032	&11.1	&Smooth\\
J112009.6-120541.1	&0.3533	&235	&\nodata	&1.45	&20.89	&0.004	&23.90	&0.030	&10.4	&Smooth\\
J112009.3-120830.4	&0.3672	&128	&\nodata	&2.13	&19.74	&0.001	&23.67	&0.028	&10.9	&Smooth\\
J112007.5-120509.1	&0.3532	&0	&\nodata	&3.20	&19.28	&0.002	&23.25	&0.020	&11.3	&Smooth\\
J112007.4-120953.7	&0.3671	&384	&\nodata	&2.20	&19.46	&0.001	&23.62	&0.024	&11.1	&Smooth\\
J112007.0-120455.7	&0.3525	&77	&\nodata	&1.39	&20.84	&0.003	&24.55	&0.038	&10.4	&Smooth\\
J112006.8-121008.1	&0.3700	&471	&\nodata	&0.42	&21.36	&0.002	&23.99	&0.029	&10.7	&Smooth\\
J112006.2-120920.0	&0.3695	&333	&\nodata	&2.37	&20.96	&0.004	&24.80	&0.056	&10.3	&Smooth\\
J112024.6-120311.1	&0.3731	&471	&\nodata	&2.17	&19.00	&0.001	&23.42	&0.017	&11.4	&Smooth\\
J112014.6-120360.0	&0.3691	&655	&\nodata	&1.73	&21.64	&0.007	&25.13	&0.051	&9.9	&Smooth\\
J112014.5-120422.7	&0.3496	&589	&0.20	&1.93	&23.32	&0.023	&24.92	&0.038	&8.3	&Smooth\\
J112029.6-120134.0	&0.3688	&568	&\nodata	&2.34	&20.55	&0.003	&24.73	&0.041	&10.6	&Smooth\\
J112024.3-120142.4	&0.3717	&164	&0.10	&3.58	&19.58	&0.002	&23.86	&0.026	&11.0	&Smooth\\
J112017.3-120211.5	&0.3675	&399	&0.47	&2.98	&20.60	&0.004	&24.17	&0.032	&10.0	&Clumpy\\
J112011.0-120831.0	&0.3680	&123	&0.03	&2.57	&19.28	&0.001	&23.46	&0.026	&11.1	&Smooth\\
J112008.0-120542.3	&0.3511	&174	&\nodata	&2.02	&19.62	&0.001	&23.18	&0.018	&10.9	&Smooth\\
J112009.2-120434.0	&0.3529	&220	&\nodata	&3.87	&20.98	&0.006	&24.57	&0.040	&10.1	&Unclassified\\
J112006.6-120937.0	&0.3642	&358	&\nodata	&1.27	&20.88	&0.002	&24.61	&0.036	&10.4	&Smooth\\
J112006.3-121027.8	&0.3742	&578	&\nodata	&2.41	&20.26	&0.002	&23.93	&0.034	&10.7	&Smooth\\
J112010.1-121012.6	&0.3695	&415	&6.10	&6.20	&18.75	&0.001	&22.52	&0.013	&11.1	&Clumpy\\
J112005.8-120443.4	&0.3514	&184	&0.35	&2.86	&19.41	&0.002	&23.35	&0.021	&11.0	&Smooth\\
J112005.1-120529.0	&0.3501	&210	&\nodata	&1.36	&20.77	&0.003	&23.94	&0.033	&10.4	&Smooth\\
J112004.3-120537.7	&0.3727	&287	&0.23	&2.57	&20.22	&0.002	&22.99	&0.017	&10.1	&Clumpy\\
J112018.8-120732.1	&0.3649	&778	&\nodata	&2.25	&20.53	&0.002	&22.59	&0.010	&9.7	&Clumpy\\
J112016.8-120906.6	&0.3660	&517	&0.37	&2.35	&21.10	&0.003	&23.81	&0.027	&9.4	&Clumpy\\
J112013.4-120747.3	&0.3714	&415	&\nodata	&1.41	&21.10	&0.003	&24.56	&0.040	&10.2	&Smooth\\
J112006.3-120845.3	&0.3724	&297	&\nodata	&1.86	&21.58	&0.004	&25.32	&0.079	&10.0	&Unclassified\\
J112020.6-120806.7	&0.3541	&839	&0.26	&1.75	&20.73	&0.002	&23.66	&0.024	&10.1	&Smooth\\
J112019.9-120817.7	&0.3327	&768	&\nodata	&1.99	&19.98	&0.002	&22.91	&0.018	&10.5	&Clumpy\\
J112015.9-120847.8	&0.3710	&445	&\nodata	&2.00	&19.71	&0.001	&23.54	&0.021	&10.8	&Smooth\\
J112012.7-120802.3	&0.3535	&322	&\nodata	&2.34	&19.98	&0.002	&24.30	&0.032	&10.8	&Smooth\\
J112010.2-120851.6	&0.3720	&0	&\nodata	&4.04	&18.56	&0.001	&22.76	&0.020	&11.7	&Smooth\\
J112008.8-120859.3	&0.3713	&113	&\nodata	&2.71	&19.65	&0.001	&23.82	&0.032	&11.0	&Smooth\\
J112007.5-120857.3	&0.3756	&210	&\nodata	&1.40	&20.32	&0.002	&24.32	&0.035	&10.6	&Smooth\\
J112009.5-120828.5	&0.3688	&128	&0.59	&3.52	&19.21	&0.001	&23.03	&0.011	&10.9	&Smooth\\
J112021.3-120154.6	&0.3725	&82	&\nodata	&1.24	&20.79	&0.003	&24.48	&0.037	&10.6	&Smooth\\
J112011.5-120440.2	&0.3509	&343	&\nodata	&1.41	&20.80	&0.003	&24.48	&0.027	&10.4	&Smooth\\
J112009.0-120513.2	&0.3510	&123	&\nodata	&2.74	&20.41	&0.003	&23.55	&0.029	&10.4	&Smooth\\
J112007.7-120514.3	&0.3540	&36	&\nodata	&2.23	&19.97	&0.002	&23.83	&0.025	&10.9	&Smooth\\
J112019.6-115920.5	&0.3707	&502	&\nodata	&1.01	&21.58	&0.004	&25.31	&0.055	&10.2	&Smooth\\
J112018.5-120050.2	&0.3758	&404	&\nodata	&1.52	&20.93	&0.004	&24.61	&0.030	&10.4	&Smooth\\
J112019.5-120803.9	&0.3730	&763	&\nodata	&1.82	&21.25	&0.004	&25.13	&0.057	&10.3	&Smooth\\
J112014.2-120859.9	&0.3664	&312	&0.69	&6.27	&19.00	&0.002	&21.93	&0.014	&10.5	&Clumpy\\
J112005.1-120935.5	&0.3717	&445	&\nodata	&1.40	&21.35	&0.003	&25.17	&0.057	&10.2	&Smooth\\
J112014.1-120415.8	&0.3700	&578	&\nodata	&2.74	&20.11	&0.003	&24.51	&0.033	&10.7	&Unclassified\\
J112021.1-120015.0	&0.3755	&471	&\nodata	&1.69	&20.05	&0.002	&23.61	&0.023	&10.7	&Smooth\\
J112013.0-120319.9	&0.3724	&701	&\nodata	&2.53	&21.83	&0.009	&24.77	&0.054	&9.7	&Unclassified\\
J112008.8-120026.0	&0.3867	&599	&\nodata	&2.77	&21.36	&0.007	&23.26	&0.022	&9.4	&Clumpy\\
J112013.3-115847.1	&0.3735	&20	&\nodata	&2.56	&19.64	&0.002	&23.68	&0.025	&11.1	&Smooth\\
J112015.3-115708.4	&0.3537	&543	&\nodata	&1.99	&18.99	&0.001	&23.29	&0.013	&11.2	&Smooth\\
J112013.3-115850.6	&0.3706	&0	&0.24	&2.58	&18.64	&0.001	&22.21	&0.004	&11.5	&Smooth\\
J112013.2-115841.4	&0.3704	&46	&\nodata	&1.79	&19.77	&0.002	&23.79	&0.023	&11.0	&Smooth\\
J112010.4-120151.9	&0.3771	&906	&2.45	&2.43	&18.97	&0.001	&20.78	&0.002	&11.0	&Smooth\\
J112010.5-120220.1	&0.3750	&896	&1.04	&4.32	&20.10	&0.004	&24.40	&0.039	&10.4	&Clumpy\\
J112008.4-120012.9	&0.3680	&568	&\nodata	&1.66	&21.04	&0.004	&24.65	&0.041	&10.3	&Smooth\\
J112014.6-115913.5	&0.3711	&154	&\nodata	&1.43	&20.63	&0.003	&24.41	&0.030	&10.6	&Smooth\\
J112015.7-115923.7	&0.3712	&251	&\nodata	&1.88	&19.79	&0.001	&23.89	&0.026	&10.9	&Smooth\\
J112016.6-120106.4	&0.3684	&476	&\nodata	&2.83	&19.48	&0.002	&23.73	&0.024	&11.0	&Smooth\\
J112016.6-120058.7	&0.3680	&497	&\nodata	&1.92	&21.52	&0.006	&24.73	&0.037	&9.3	&Clumpy\\
J112014.4-120114.0	&0.3690	&619	&0.70	&4.26	&19.35	&0.002	&21.98	&0.011	&10.2	&Clumpy\\
J112013.8-115912.2	&0.3705	&118	&\nodata	&2.04	&20.07	&0.002	&24.22	&0.032	&10.9	&Smooth\\
J112012.7-115955.0	&0.3706	&333	&0.13	&1.99	&21.15	&0.004	&23.16	&0.015	&9.4	&Clumpy\\
J112015.5-120143.9	&0.3753	&512	&0.16	&4.77	&19.26	&0.002	&24.20	&0.033	&11.0	&Smooth\\
J112017.8-120024.8	&0.3744	&537	&\nodata	&3.02	&21.13	&0.006	&23.21	&0.022	&9.4	&Smooth\\
J112017.9-115942.3	&0.3690	&435	&\nodata	&1.03	&21.70	&0.005	&25.47	&0.056	&10.1	&Smooth\\
J112008.9-120931.1	&0.3721	&225	&\nodata	&2.45	&20.04	&0.002	&23.89	&0.024	&10.8	&Smooth\\
J112008.9-120819.7	&0.3683	&189	&\nodata	&1.62	&20.34	&0.002	&23.83	&0.029	&10.5	&Smooth\\
J112009.4-120906.4	&0.3673	&97	&\nodata	&2.62	&20.09	&0.001	&23.76	&0.032	&10.7	&Smooth\\
J112009.7-120920.0	&0.3652	&148	&\nodata	&1.51	&20.40	&0.002	&24.35	&0.038	&10.6	&Smooth\\
J112010.7-121104.0	&0.3500	&681	&0.53	&4.63	&19.86	&0.002	&23.98	&0.031	&10.6	&Clumpy\\
J112010.0-120601.5	&0.3580	&333	&\nodata	&1.72	&21.86	&0.005	&24.62	&0.040	&9.7	&Clumpy\\
J112011.0-121108.5	&0.3510	&701	&\nodata	&2.59	&20.63	&0.003	&22.74	&0.022	&9.5	&Clumpy\\
J112010.9-120836.3	&0.3680	&97	&\nodata	&0.86	&22.14	&0.005	&24.88	&0.066	&10.0	&Smooth\\
J112011.5-120928.7	&0.3692	&215	&\nodata	&2.65	&20.94	&0.003	&24.91	&0.040	&10.3	&Smooth\\
J112011.5-120828.3	&0.3702	&159	&\nodata	&1.90	&20.99	&0.003	&24.28	&0.040	&10.1	&Smooth\\
J112011.3-121114.2	&0.3667	&737	&\nodata	&2.42	&22.06	&0.009	&23.80	&0.031	&9.0	&Clumpy\\
J112015.1-120722.9	&0.3720	&589	&0.17	&3.43	&20.90	&0.004	&24.85	&0.057	&10.3	&Unclassified\\
J112014.8-120742.3	&0.3692	&507	&\nodata	&2.35	&20.75	&0.003	&24.07	&0.038	&10.4	&Unclassified\\
J112015.1-120746.3	&0.3720	&507	&\nodata	&2.24	&20.53	&0.003	&24.60	&0.043	&10.6	&Smooth\\
J112015.9-120712.7	&0.3706	&671	&\nodata	&2.23	&19.07	&0.001	&23.13	&0.016	&11.2	&Smooth\\
J112016.5-120812.8	&0.3713	&527	&\nodata	&2.30	&20.63	&0.003	&24.09	&0.037	&10.4	&Smooth\\
J112017.1-120940.2	&0.3512	&589	&0.46	&4.90	&19.67	&0.002	&24.26	&0.019	&9.9	&Clumpy\\
J112017.9-120554.8	&0.3900	&834	&\nodata	&3.15	&22.19	&0.012	&23.89	&0.037	&9.0	&Unclassified\\
J112019.5-120536.0	&0.3746	&932	&1.27	&4.98	&19.22	&0.001	&23.88	&0.032	&10.6	&Clumpy\\
J112020.0-120805.8	&0.3449	&793	&\nodata	&2.27	&22.14	&0.008	&24.18	&0.029	&9.0	&Clumpy\\
J112010.9-120234.9	&0.3744	&829	&0.61	&5.16	&20.10	&0.004	&22.67	&0.017	&9.9	&Clumpy\\
J112009.6-120114.1	&0.3698	&788	&\nodata	&2.14	&20.16	&0.003	&24.17	&0.032	&10.7	&Smooth\\
J112010.9-115752.1	&0.3692	&353	&\nodata	&3.06	&19.02	&0.001	&23.28	&0.019	&11.4	&Smooth\\
J112023.3-120329.7	&0.3758	&537	&\nodata	&2.89	&21.14	&0.005	&24.47	&0.044	&9.8	&Clumpy\\
J112010.5-120223.8	&0.3744	&880	&\nodata	&3.07	&21.59	&0.011	&24.33	&0.038	&9.3	&Clumpy\\
J112010.9-120207.8	&0.3770	&875	&\nodata	&3.98	&20.73	&0.005	&22.61	&0.012	&9.6	&Clumpy\\
J112010.9-120315.3	&0.3497	&640	&0.21	&2.68	&20.32	&0.003	&21.81	&0.009	&9.7	&Clumpy\tablenotemark{f }\\
J112011.5-115935.2	&0.3681	&266	&\nodata	&0.84	&21.95	&0.006	&25.50	&0.064	&10.0	&Smooth\\
J112012.4-120250.3	&0.3747	&804	&0.17	&3.43	&20.51	&0.004	&24.65	&0.048	&10.5	&Unclassified\\
J112012.6-115910.6	&0.3712	&118	&\nodata	&1.89	&20.24	&0.002	&23.97	&0.032	&10.7	&Clumpy\\
J112013.2-120345.2	&0.3684	&614	&0.62	&6.39	&19.47	&0.003	&23.04	&0.015	&10.1&Clumpy\tablenotemark{f }\\
J112014.0-115744.5	&0.3339	&343	&0.07	&2.82	&19.93	&0.003	&23.20	&0.016	&10.7	&Smooth\\
J112014.0-115913.1	&0.3692	&128	&\nodata	&2.02	&20.17	&0.002	&24.02	&0.032	&10.8	&Smooth\\
J112014.1-115814.3	&0.3325	&195	&\nodata	&3.72	&21.14	&0.007	&22.70	&0.019	&9.4	&Clumpy\\
J112014.5-115808.9	&0.3742	&230	&0.19	&3.47	&20.11	&0.003	&23.77	&0.022	&9.9	&Clumpy\tablenotemark{f }\\
J112014.3-115739.8	&0.3713	&369	&\nodata	&3.76	&21.60	&0.011	&23.83	&0.036	&9.4	&Smooth\\
J112014.7-120117.2	&0.3701	&594	&\nodata	&1.77	&20.89	&0.004	&24.21	&0.038	&10.3	&Smooth\\
J112014.6-115718.1	&0.3741	&481	&0.27	&4.27	&20.05	&0.003	&22.19	&0.013	&9.9	&Clumpy\\
J112014.9-115825.6	&0.3736	&174	&0.12	&3.14	&21.00	&0.004	&23.89	&0.030	&9.7	&Clumpy\\
J112015.9-115855.1	&0.3692	&200	&\nodata	&1.79	&19.73	&0.001	&23.49	&0.021	&10.9	&Smooth\\
J112016.8-120156.3	&0.3712	&420	&\nodata	&2.41	&19.44	&0.002	&23.18	&0.019	&11.0	&Smooth\\
J112017.3-120132.4	&0.3711	&384	&\nodata	&1.81	&19.65	&0.002	&23.63	&0.022	&11.0	&Smooth\\
J112017.6-120241.1	&0.3338	&450	&1.04	&4.53	&20.29	&0.003	&22.42	&0.016	&9.7	&Clumpy\\
J112017.7-115757.2	&0.3549	&435	&0.39	&2.86	&20.39	&0.003	&23.24	&0.026	&9.7	&Clumpy\\
J112018.2-120256.8	&0.3761	&476	&0.10	&6.56	&20.23	&0.005	&23.01	&0.019	&9.9	&Smooth\\
J112017.8-120230.5	&0.3708	&410	&\nodata	&1.55	&20.84	&0.004	&24.82	&0.043	&10.5	&Smooth\\
J112018.5-120324.3	&0.3737	&578	&\nodata	&4.41	&19.91	&0.003	&23.92	&0.030	&10.8	&Smooth\\
J112019.2-120230.7	&0.3701	&328	&\nodata	&1.89	&19.73	&0.002	&23.62	&0.020	&10.9	&Smooth\\
J112020.1-120017.2	&0.3692	&481	&0.19	&3.70	&20.12	&0.003	&22.43	&0.019	&9.9	&Clumpy\\
J112021.6-120143.2	&0.3716	&46	&\nodata	&1.37	&20.66	&0.003	&24.68	&0.031	&10.7	&Smooth\\
J112021.1-120135.5	&0.3677	&97	&\nodata	&2.84	&21.42	&0.007	&22.74	&0.019	&9.4	& Clumpy\tablenotemark{f }\\
J112021.3-120215.6	&0.3735	&169	&\nodata	&2.48	&20.43	&0.003	&24.76	&0.048	&10.7	&Smooth\\
J112021.8-120208.2	&0.3712	&118	&\nodata	&1.67	&20.86	&0.004	&24.57	&0.041	&10.5	&Smooth\\
J112022.7-120307.5	&0.3742	&420	&\nodata	&3.82	&18.66	&0.001	&23.19	&0.020	&11.5	&Smooth\\
J112021.7-120250.4	&0.3686	&333	&\nodata	&1.64	&21.02	&0.003	&24.70	&0.046	&10.4	&Smooth\\
J112022.6-120153.2	&0.3455	&46	&0.90	&2.45	&20.21	&0.002	&22.61	&0.016	&9.9	& Clumpy \tablenotemark{f }\\
J112022.9-120151.4	&0.3714	&56	&\nodata	&0.99	&20.30	&0.002	&24.29	&0.028	&10.8	&Smooth\\
J112022.7-120141.4	&0.3708	&46	&\nodata	&1.10	&21.04	&0.003	&24.94	&0.037	&10.4	&Smooth\\
J112023.2-120127.6	&0.3731	&123	&\nodata	&1.26	&21.29	&0.004	&25.38	&0.059	&10.2	&Smooth\\
J112025.0-120324.1	&0.3723	&548	&\nodata	&2.35	&19.39	&0.001	&23.53	&0.022	&11.2	&Smooth\\
J112024.2-120316.1	&0.3703	&486	&\nodata	&1.61	&20.00	&0.001	&23.71	&0.026	&10.9	&Smooth
\enddata
\tablenotetext{a}{Names follow the SDSS format: JHHMMSS.s$\pm$DDMMSS.s for R.A. and decl. in sexagesimal units. }
\tablenotetext{b}{Errors corresponding to adopting different conversion factors for F$_{24\mu m}$ to F$_{8-1000\mu m}$ are $\sim10-20$\% .}
\tablenotetext{c}{For galaxies with \rhalf$>1$~kpc, Source Extractor measurements of the error on their spheroid radii are $\sim 1-8$\% .}
\tablenotetext{d}{Magnitudes are observed-frame.}
\tablenotetext{e}{See \citet{Tran2009} for errors in mass calculation. }
\tablenotetext{f}{Jellyfish candidate}
\end{deluxetable*}